# Kidney segmentation using 3D U-Net localized with Expectation Maximization

Omid Bazgir, Kai Barck, Richard A.D. Carano, Robby M. Weimer, Luke Xie*

Genentech, South San Francisco, California, USA

Luke.xie@gene.com

*Abstract—* **Kidney volume is greatly affected in several renal diseases. Precise and automatic segmentation of the kidney can help determine kidney size and evaluate renal function. Fully convolutional neural networks have been used to segment organs from large biomedical 3D images. While these networks demonstrate state-of-the-art segmentation performances, they do not immediately translate to small foreground objects, small sample sizes, and anisotropic resolution in MRI datasets. In this paper we propose a new framework to address some of the challenges for segmenting 3D MRI. These methods were implemented on preclinical MRI for segmenting kidneys in an animal model of lupus nephritis. Our implementation strategy is twofold: 1) to utilize additional MRI diffusion images to detect the general kidney area, and 2) to reduce the 3D U-Net kernels to handle small sample sizes. Using this approach, a Dice similarity coefficient of 0.88 was achieved with a limited dataset of n=196. This segmentation strategy with careful optimization can be applied to various renal injuries or other organ systems.**

*Keywords-Magnetic Resonance Image, Kidney Segmentation, Localization, Deep Learning, Convolutional Neural Network*

## I. Introduction

Kidney function and activity is highly dependent on kidney volume in a variety of diseases such as polycystic kidney disease, lupus nephritis, renal parenchymal disease, and kidney graft rejection [1]. Automatic evaluation of the kidney through imaging has the potential to accurately stratify patients and determine outcome. *In vivo* imaging modalities offer unique strengths and limitations. MRI, in particular, does not have ionizing radiation, is not operator dependent, and has good tissue contrast that enables kidney segmentation and volume related information. Traditional methods have been used to evaluate the kidney more locally, such as manual tracing, stereology, or general image processing. These methods can be labor intensive or inconsistent [1, 2]. To address these issues, we propose to use an integrated deep learning model to segment the kidney.

Deep learning segmentation frameworks can be used to automatically decipher the kidney in volumetric MRI datasets, as they outperform traditional models including [3]. Three dimensional convolutional neural networks (CNNs) have been trained end-to-end to delineate objects of interest, which contains coupled convolutional and deconvolutional layers such as V-Net [4] and 3D U-Net [5]. Although 3D CNNs offer state-of-the-art performance, they suffer from high computational cost and memory consumption, which limits their field-of-view and depth [6]. Hence, these networks can be particularly problematic for segmenting small objects in limited images typically found in MRI studies. MRI tends to include a large field-of-view or background for preventing aliasing artifacts. When the background represents a significant portion, the network may not be optimally trained to segment the foreground object of interest. This can be the case even with a weighted loss function [7]. Thus, an alternative strategy is needed to reduce the parameters of a large 3D segmentation network, avoid overfitting, and improve network performance.

First, to address the issue of the background effect, we incorporated a derived MRI contrast mechanism for the localization step prior to learned segmentation. Second, we modified the 3D U-Net to reduce the number of parameters and incorporated a Dice loss function for the segmentation. Third, we incorporated augmentation and MRI histogram matching to increase the number of training datasets. We also applied our method on super resolved images of our dataset to determine whether enhanced images can improve segmentation performance. These methods were implemented on preclinical MRI using an animal model of lupus nephritis.

## II. Methodology

### A. Animal model and data acquisition

Fifteen friend Virus B female mice were used for this study, where 8 were used for the lupus nephritis (LN) disease group and 7 for the control group (The Jackson Laboratory, Sacramento, CA). Animals were imaged every 2 weeks for 4 timepoints starting at 13 weeks of age. At each time point, multiple MRI datasets were acquired for each animal. A total of 196 3D MR images were acquired for this study. All images were manually segmented by a single user. Kidneys were outlined slice by slice for the entire image volume using Amira (Thermo Fisher Scientific, Hillsboro, OR). During MR imaging, animals were anesthetized under isoflurane, breathing freely, and maintained at 37 C. MRI was performed on a Bruker 7T (Billerica, MA) with a volume transmit and cryogenic surface receive coil. A custom in vivo holder was constructed with 3D printing (Stratasys Dimension) to provide secure positioning of the brain and spine. MRI diffusion tensor imaging was performed (single-shot EPI) with individual local shims using the following parameters: TR=4 s, TE=42 ms, BW=250 kHz, diffusion directions=12, FOV=22×22 mm$^2$, encoding matrix=110×110, slices=15, image resolution=200×200 μm$^2$, slice thickness=1 mm, acquisition time=13 min. Diffusion tensor parametric maps were computed, which include: fractional anisotropy (FA), mean diffusivity (MD), axial diffusivity (AD), and radial diffusivity (RD). FA and MD images were used for the integrated semantic segmentation algorithm. Animal procedures were approved by the Genentech Institutional Animal Care and Use Committee.

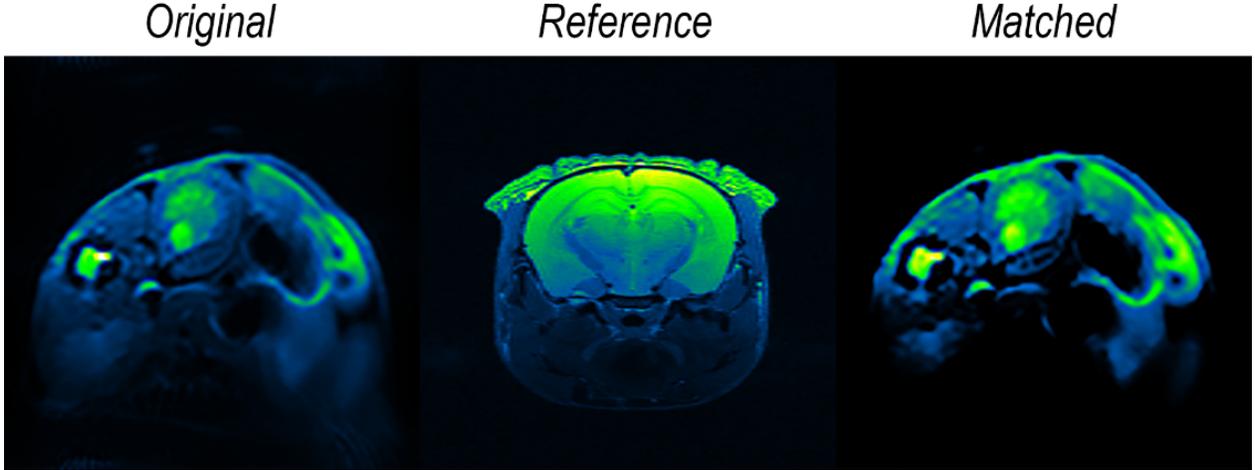

*Fig 1. Histogram matching was used to simulate other contrasts and increase variance of the training dataset. Each image of the training set (left image) was histogram matched with reference image (center image) to generate new set of images (right image).*

### B. Localization with Expectation Maximization (EM)

The FA images were used for the localization step. They were segmented using EM, which was initialized with K-means (12 classes) heuristically [8]. The general kidney vicinity was isolated using one of the tissue classes and used as the detected object. These parameters were used for the algorithm: number of iterations for convergence=7 and Markov random field smoothing factor=0.05.

### C. Data Agumentation

MD images were histogram matched with a mouse brain dataset to generate new datasets (Fig. 1). Both datasets were rotated 90°, flipped left-to-right, and flipped up-and-down. Data augmentation was done only for training set, to make sure the network is validated on a completely unseen data. The total number of acquired datasets was n=196. With augmentation, the training dataset increased from n=180 to n=1800, leaving the test dataset to n=16. Note that the training and testing split was done animal-wise, where each time one animal was kept out for testing and rest was used for training.

### D. Deep semantic segmentation

The metric used for evaluating segmentation performance was the Dice similarity coefficient (DSC, Equation 1). Therefore, to train a CNN with the objective of maximizing the DSC, we minimized the DSC for all the images (Equation 2, N=number of images). Also, due to unbalanced distribution of background and kidney in the volumetric images, we used a weighted loss function [7], which we will refer to as the Dice loss (Equation 3, where $p_i$ and $q_i$ are predicted and ground truth masks, respectively).

$$\frac{2\sum_{i=1}^{N} p_i q_i}{\sum_{i=1}^{N} p_i^2 + \sum_{i=1}^{N} q_i^2} \quad (1)$$

$$Dice\ loss = 1 - \frac{2\sum_{l=1}^{2} w_l \sum_{i=1}^{N} p_{li} q_{li}}{\sum_{i=1}^{N} p_{li}^2 + \sum_{l=1}^{2} w_l \sum_{i=1}^{N} q_{li}^2} \quad (2)$$

$$w_l = \frac{1}{(\sum_{i=1}^{N} q_{li})^2} \quad (3)$$

To alleviate the background effect, we projected the EM segmentation mask in the slice direction. The boundaries of the projected 2D mask was used to define a rectangular box for object detection. The defined box was enlarged by 5 pixels on all sides to ensure coverage of the kidney. The 3D network was trained and tested on the MD images inside the detected area. The same detected area was used for the super-resolved images. Since the cropped object has arbitrary size in the first two dimensions based on the 2D projected mask, all cropped images were resized to 64×64×16 for the original resolution images and resized to 64×64×64 for the super-resolved images.

### E. Super resolution

MD images were super resolved in the through-plane direction to improve spatial resolution. The original matrix of 110×110×15 were resolved 5× to give a resultant matrix of 110×110×75. Images were enhanced using a deep super resolution neural network [9].

### III. RESULTS

Fig. 2A shows the six elements of the diffusion tensor. The changing diffusion contrast is most noticeable in the inner and outer medullary regions. The changing contrast is noticeable in the diagonal ($D_{xx}$, $D_{yy}$, $D_{zz}$) and off-diagonal elements ($D_{xy}$, $D_{xz}$, $D_{yz}$). Consequently, the contrast does not change in the cortex, thus resulting in a very low FA (Fig. 2B). This low FA allowed the kidney to be segmented from the background. MR images were super resolved in the through-plane direction as shown in Fig. 2C. The improvements are most obvious in the sagittal and coronal directions. In-plane resolution is minimally affected as shown in the axial slice (Fig. 2C). Fig. 3A shows the results of training a 3D U-Net on MD images, without any preprocessing. The DSC plot shows a uniform distribution with a mean of 0.49. In Fig. 3B the abdominal area is detected as foreground with connected component analysis and cropped using the MD images. The DSC plot displays a normal distribution with a mean of 0.52. Fig. 3C shows the results using EM segmentation alone. A mean DSC of 0.65 was achieved. Fig. 3D represents the results of the integrated strategy: first the kidney was detected using EM segmentation on FA images, then 3D U-Net was trained on the detected kidney area from MD images. The average DSC of this approach was 0.88. The DSC plot of semantic segmentation with super-resolved MD images (Fig. 3E) is very similar to semantic segmentation at the original resolution (Fig. 3D). Here, the average DSC was 0.86. The results are summarized in Table 1 with additional comparison metrics, such as volume difference (VD) and positive predictive value (PPV).

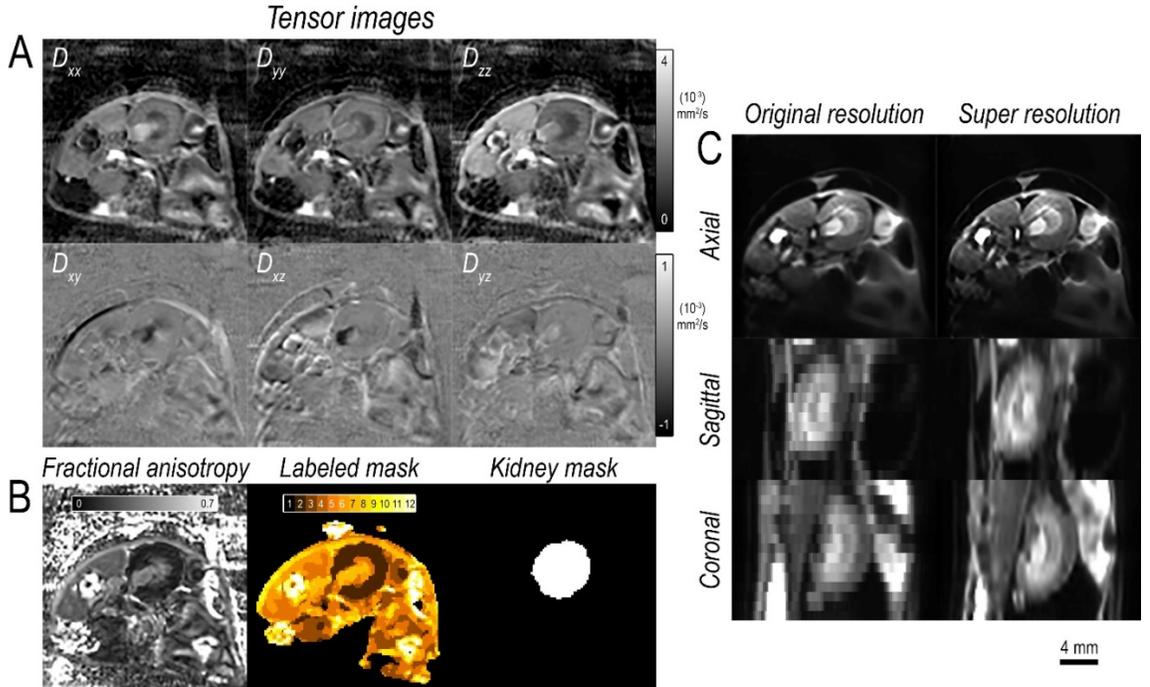

*Fig 2. A: Diffusion tensor elements. B: Fractional anisotropy image used for EM segmentation (12 classes) and object detection steps. C: Super resolved images in the slice direction. Scale bar = 4 mm.*

*Table 1. mean and standard deviation of segmentation results using DSC, VD, and PPV. The best value for each method is shown in bold.*

| Method | DSC | VD | PPV |
|---|---|---|---|
| 3D U-Net | 0.49 ± 0.28 | 0.36 ± 0.2 | 0.67 ± 0.08 |
| CC + 3D U-Net | 0.52 ± 0.21 | 0.31 ± 0.17 | 0.71 ± 0.07 |
| EM | 0.65 ± 0.23 | 0.16 ± 0.15 | 0.76 ± 0.07 |
| **Proposed** | **0.88 ± 0.10** | **0.09 ± 0.05** | **0.94 ± 0.05** |
| Proposed + SR | 0.86 ± 0.12 | 0.08 ± 0.05 | 0.93 ± 0.05 |

## IV. DISCUSSION AND CONCLUSION

This work demonstrates the integration of EM based localization and 3D U-Net for kidney segmentation. The localization step led to a significantly improved result of the deep learning method. We also demonstrated that while the EM segmentation was critical for the performance of deep learning, this segmentation method alone performed poorly. EM segmentation method isolated the kidney in the central slice, however, it did not preserve the joint representation of kidney volume. Thus, the central slice was used for all slices across the volume as the detected rectangular object. A weighted Dice loss can be crucial for the error minimization and balance of the object and background. Without the localization step, we found that the performance did not significantly increase with the inclusion of a weighted Dice loss. Consequently, the background contained objects and organs that appeared similar to the kidney. Note that, due to scarcity of existing pretrained models, incorporating transfer learning methods for can be challenging. Instead of using transfer learning we focused on training a problem specific

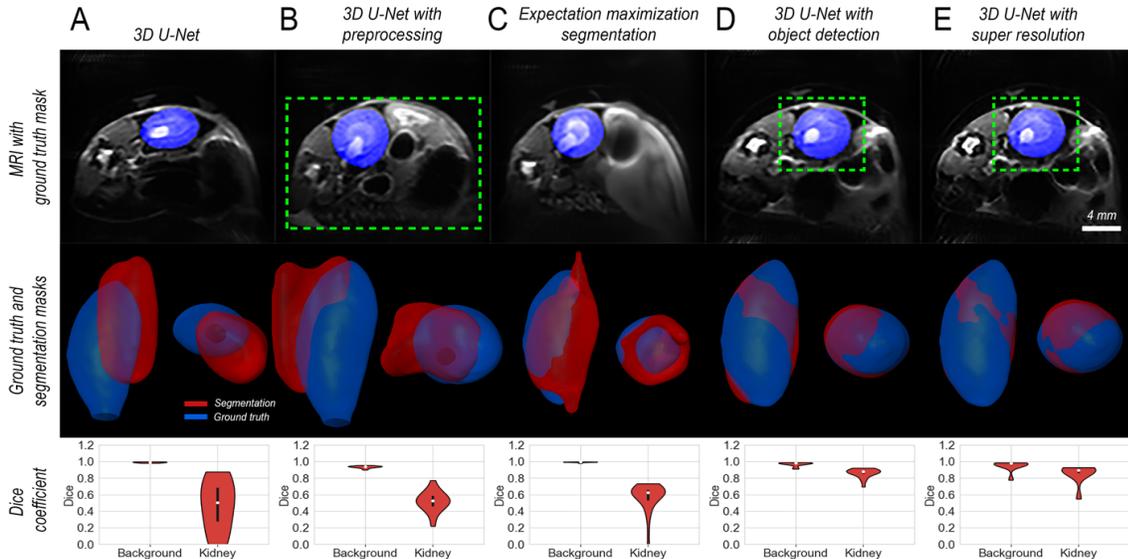

Fig 3. Segmentation results using various strategies. A: 3D U-Net. B: detecting the foreground with connected component preprocessing. C: EM segmentation. D: kidney detection via EM segmentation. E: kidney detection via EM segmentation on super-resolved images. First row: ground truth manual labels overlaid on MRI. Second row: transparent surface renderings of the ground truth and segmentation masks. Coronal and axial views are shown in pairs. Third row: DSCs shown as violin plots. Example datasets were selected based on the mean DSCs for each segmentation strategy. All segmentation results are 3D U-Net based except for C, which is only EM segmentation. Green box indicates the area for object detection. Scale bar=4m

neural network. In particular, the localization of the kidney was a critical step in developing an overall accurate segmentation method. Most likely, using an existing network with transfer learning not designed for this problem would have resulted in lower accuracy.

In the super-resolved images, the DSC was slightly lower compared with the original resolution images. This may be due to the fact that manual labeling of the kidneys was performed on the original low-resolution images. One solution would be to re-label all the images in the super-resolved images. The challenge here would be that a user needs to label 75 slices as opposed to the original 15 slices.

The approach presented in this study reduced the background effect and decreased the complexity of the data. Consequently, the complexity of the network can be decreased by reducing the number of kernels per convolutional layer by one half. In present study, a DSC of 0.88 was achieved with a limited MRI dataset of n=196.

In conclusion, localization is crucial for segmenting the kidney using a 3D U-net framework. The network can be used to automatically and accurately segment the kidney without the need of manual tracing. The isolation of the kidney and measurement of kidney volume can be used to evaluate renal function in diseases that greatly affect kidney size. This strategy with careful optimization can be translated for various segmentation applications.